# A Monitoring Method for the Ice Shape and the Freeze-Thaw Process of Ice Accretion on Transmission Lines Based on Circular FBG Plane Principal Strain Sensor

Zhuoke Qin, Bin Jia, Xiahui Shen, Lizhen Zhang, Honggang Lu, Chao Du, Liqin Cui, Li Zhang and Xiao Deng

*Abstract*—As a key infrastructure for China's "West-to-East Power Transmission" project, transmission lines (TL) face the threat of ice accretion under complex microclimatic conditions. This study proposes a plane principal strain sensing method based on a fiber Bragg grating circular array, achieving synchronous monitoring of 6 strains (ranging from -2000 to 2000 με) across the TL cross-section. Through finite element simulation experiments, a mapping relationship between the bending of TL and the plane principal strain has been established. After completing the sensor calibration, an experimental platform for the freeze-thaw process of ice accretion on the TL was built. The relationships between ice mass and bending strain, as well as the ice shape on the TL cross-section (C-shaped and circular ice) and plane principal strain, were studied. Furthermore, a BP neural network model was developed to determine the 4 states of the icing process (no ice/freeze/stable/thaw), achieving an accuracy of 91.23%. This study provides effective monitoring of the freeze-thaw process of ice accretion on the TL, offering important technical support for the prevention and control of ice accretion in power grid.

This work was supported in part by the National Natural Science Foundation of China under Grant 62375198 and Grant 62203320, Fundamental Research Program of Shanxi Province under Grant 202303021221028, Shanxi Scholarship Council of China under Grant 2023-039, and the Science and Technology Major Special Project of Shanxi Province under Grant 202201010101005. (*Corresponding author: Bin Jia, Xiao Deng*).

Z. Qin, B. Jia, X. Shen, H. Lu, C. Du, L. Cui and L. Zhang are with the College of Physics and Optoelectronics, Taiyuan University of Technology, Taiyuan 030024, China (e-mail: qinzhuoke1170@link.tyut.edu.cn; jiabin@tyut.edu.cn; 2024522038@link.tyut.edu.cn; 2023521655@link.tyut.edu.cn; duchao@tyut.edu.cn; cuiliqin@tyut.edu.cn; zhangli@tyut.edu.cn).

L. Zhang is with the Scientific Research Office of Shanxi Electric Power Technical College. Taiyuan 030021, China (e-mail: lixin01@tyut.edu.cn).

X. Deng is with the College of Physics and Optoelectronics, and the Key Laboratory of Advanced Transducers and Intelligent Control System, Ministry of Education, and the Shanxi Key Laboratory of Precision Measurement Physics, Taiyuan University of Technology, Taiyuan 030024, China (e-mail: dengxiao@tyut.edu.cn).

## I. INTRODUCTION

IN pursuit of achieving "Dual Carbon" goals of "Carbon Peak" and "Carbon Neutrality", China is vigorously developing clean energy [1]. The "West-to-East Power Transmission" project is a major initiative in China to optimize the distribution of energy resources. It aims to transport abundant hydro, wind, and solar power from the western regions to the economically developed, high-demand areas in the east via the power grid [2]. During the interregional transfer of energy, transmission lines (TL) pass through various complex microclimate environments. In winter and spring, cold waves may lead to localized ice accretion on the TL, presenting significant risks to their safety and operational reliability [3]. A comprehensive de-icing strategy that considers multiple factors can effectively address the ice accretion hazards on the TL, with the freeze-thaw process being a key factor influencing the de-icing measures for TL [4], [5]. Therefore, effectively monitoring the freeze-thaw process of ice accretion on the TL under local microclimates is essential for ensuring the reliable operation of large-scale power grid.

Currently, the methods for monitoring ice accretion on the TL are mainly divided into two categories: non-contact and contact-based approaches. Non-contact measurement methods mainly include manual observation and image recognition. The manual observation typically involves visually inspecting and weighing collected ice samples to obtain information such as the ice shape [6]. Huang *et al*. used two cameras to capture pictures of TL, and analyzing the shape and thickness of ice accretion on the TL [7]. Li *et al*. applied support vector regression to optimize Canny parameters, improving edge detection in ice accretion images and facilitating ice thickness identification in complex image backgrounds [8]. However, these methods are unable to detect the internal changes occurring during the freeze-thaw process of ice accretion. In addition, under harsh environmental conditions, low observation efficiency and challenges such as lens obstruction by ice, hinder the real-time monitoring of ice accretion on the TL [9].

Contact methods usually place sensors directly on the TL to monitor ice accretion, and primarily include electrical sensors and fiber-optic sensors. Electrical sensors measure force [10], inclination [11], capacitance [12], [13], vibration [14], and ultrasound [15]. It can convert properties such as the weight, bending, dielectric constant, and natural frequency of TL after icing into real-time electrical signals, thereby reflecting the ice coating conditions. However, the strong magnetic fields around TL can cause significant electromagnetic interference, compromising the accuracy of measurements taken by electrical sensors [16].

Optical fiber sensing technology, with its high precision and immunity to electromagnetic interference [17], has shown great promise for ice accretion detection on the TL. Currently, the two main technical approaches used are the distributed optical fiber method and the fiber Bragg grating (FBG) method. Sun *et al.* proposed a distributed optical fiber method for detecting ice accretion (ice thickness) based on the temperature difference between iced and non-iced TL sections, using Brillouin optical time domain reflectometry (BOTDR) for temperature measurements [18]. Xu *et al.* embedded optical fibers in the TL to measure internal temperature and identified a phase difference between the measured temperature curve and the ambient temperature curve, proposed a criterion for ice formation and developed a formula for estimating ice thickness [19]. Based on the phase-sensitive optical time-domain reflectometer (Φ-OTDR) distributed sensing technology, Ding *et al.* measured the axial strain of the TL to obtain the ice thickness [20]. Although distributed optical fiber sensing methods offering potential for long-distance ice accretion detection on the TL, they still have the limitations of local fine deployment, which is crucial for optimizing de-icing strategies (such as melting time and current) [7], [21].

FBG have advantages of small size, high precision and ease of deployment, making them ideal for the fine measurement of ice accretion on the TL. Corresponding sensors for strain, weight, and other parameters have already been developed. Luo *et al.* utilized FBG strain sensors to measure the axial local deformation of TL under ice load, and using the elastic catenary model, inferred the TL's bending conditions due to ice accretion [22]. Zhang *et al.* used a silica-based FBG strain sensor to weigh the determinand lines near TL. Based on the relationship between the weight and ice thickness, they obtained ice thickness values in the range of 0 to 30 mm [23]. Ma *et al.* designed a weight sensor with variable-sensitivity using two FBGs, which was placed between the transmission tower and insulator string to measure ice thickness through weight changes [24]. Additionally, combining FBG sensing with intelligent algorithms can further enhance the accuracy and real-time capabilities of ice accretion monitoring on the TL. By incorporating microclimate data such as temperature, humidity, and wind speed into predictive models like Gaussian Process Regression (GPR) [25], Artificial Neural Networks (ANN) [16], and Long Short-Term Memory (LSTM) [3], it is possible to forecast ice thickness on the TL over time.

However, accurately preventing and controlling ice accretion-related disasters on the TL requires comprehensive state data of both ice formation and thaw processes [4], [5]. At the same time, under the influence of microclimates, different shapes of ice accretion (such as C-shaped and circular ice) can form on the surface of TL, which also affects the efficiency of de-icing [7], [21]. Studies suggest that differences in structural curvature and ice shape can lead to complex deformations of TL cross-section. These complex deformations can be characterized by the plane principal strain [26]. Currently, existing optical fiber sensing methods are limited by sensor structure and deployment methods, making it impossible to simultaneously and effectively measure both the ice shape and the freeze-thaw state of ice accretion. These limitations not only restrict the accuracy of icing prediction models for TL but also prevent them from providing more comprehensive data support for de-icing strategies.

In this paper, based on the principle of plane principal strain, we propose a novel circular FBG plane principal strain sensor (CL-FBG Sensor) for monitoring ice accretion on the TL. The sensor consists of 6 sets of temperature-strain sensing units, and it can monitor TL bending, ice accretion weight, and the freeze-thaw process by measuring the plane principal strain of the cross-section. Based on finite element model (FEM) simulations, the relationship between the bending degree (direction and magnitude) of TL and the response of plane principal strain was studied. Through weight loading experiments, the relationship between the weight change caused by ice accretion and the surface deformation of TL was investigated. The freeze-thaw experiment platform for ice accretion was built to verify the sensor's performance in identifying ice shape and measuring ice mass. Meanwhile, a icing state judgment model of TL based on BP neural network was established, and the relationship between the freeze-thaw process of ice accretion and the measurement results at each stage of the sensor was theoretically analyzed.

## II. SENSOR DESIGN AND PRINCIPLE

### A. Structure Design

The CL-FBG Sensor is composed of strain units, temperature units, and circular lock. The detailed design of the proposed sensor is shown in Fig 1. 6 FBGs serve as the strain sensing units of the sensor, being sensitive to both temperature and strain. The other 6 FBGs function as temperature sensing units, providing temperature compensation for the sensor. The circular lock securely fixes the strain and temperature sensing units in a circular array on the surface of TL. The sensor can measure strain and temperature in 6 directions throughout the freeze-thaw process of ice accretion on the TL, allowing further determination of the plane principal strain of TL cross-section.

The packaging structure of the FBG is shown in Fig 1(a). A strain sensing unit (FBG-$\varepsilon$) and a temperature sensing unit (FBG-$T$) are located on the same optical fiber. To ensure that the FBG closely adheres to the surface of TL and avoids relative slipping caused by deformation with the ice, an annular sector dumbbell structure is designed at both ends of the FBG-$\varepsilon$. This dumbbell structure not only fixes the FBG to the TL surface but also increases the contact area of the optical fiber embedded in the ice. FBG-$T$ is encapsulated in a thin-walled stainless steel tube to ensure that the FBG is only affected by temperature. The encapsulated FBG-$T$ is used as a high-sensitivity temperature sensor for temperature compensation.

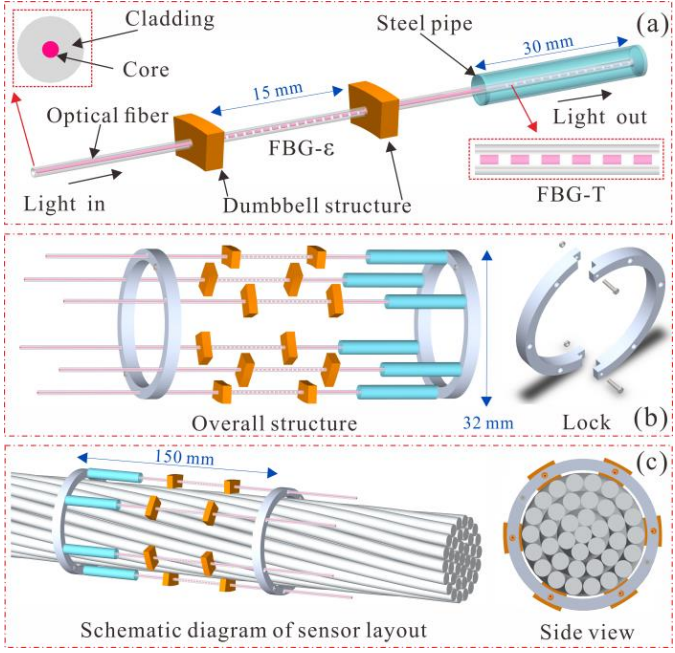

Fig 1. Sensor design. (a) FBG packaging structure. (b) Spatial structure of the sensor. (c) Sensor layout diagram.

FBG-$\varepsilon_{\#i}$ and FBG-$T_{\#i}$ ($i$=1~6) are located at the same horizontal position on the TL, and their temperatures can be considered equal. The distance between each FBG-$\varepsilon$ and FBG-$T$ is 50 mm. The FBG grating length used for strain sensing is 10 mm, and the inner diameter, outer diameter, central angle, and thickness of each annular sector dumbbell structure are 14 mm, 16 mm, 8°, and 5 mm, respectively. The FBG grating length used for temperature sensing is 5 mm, and the packaging structure length is 25 mm. The circular lock is composed of two identical semicircular rings, which are fixed using screws through the holes on both sides of the semicircles. The inner diameter of the circular lock is 14 mm, the outer diameter is 16 mm, and the width is 5 mm. On the circular lock, 6 small holes with a radius of 2 mm are evenly distributed, ensuring that the optical fiber can pass through. From the cross-sectional view, the angle between the lines connecting the small holes and the center of the circle is 60°. The distance between the two circular locks is 150 mm, as shown in Fig. 1 (b~c).

The physical image of the CL-FBG Sensor is shown in Fig. 2. The annular sector-shaped dumbbell structure and circular lock are made using nylon material with hybrid carbon fiber through 3D printing. The FBG-$T$ is placed in the middle of the capillary stainless steel to keep it in a free state. Both ends of the stainless tube is sealed with epoxy resin. The annular sector structure is fixed at both ends of the FBG-$\varepsilon$ using a heat-shrinking process. The optical fiber is passed through the hole in the right circular lock, with one end of the FBG-$T$ placed in the hole of the left circular lock. The FBG-$\varepsilon$ is pre-stressed and mounted on the surface of TL (LGJ-400/35).

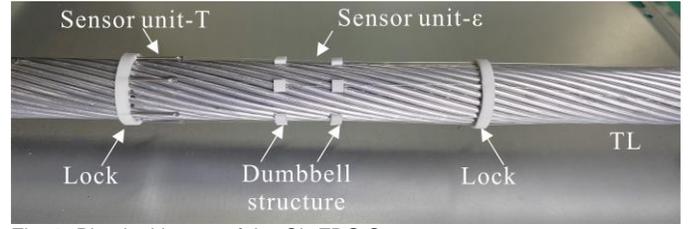

Fig. 2. Physical image of the CL-FBG Sensor.

### B. Theoretical Modeling

When the FBG is affected by mechanical strain or temperature changes, the Bragg center reflection wavelength $\lambda_B$ will shift. The influence of axial strain and temperature change on $\lambda_B$ is expressed as follows [27]:

$$\Delta\lambda_B = K_\varepsilon \Delta\varepsilon + K_T \Delta T \quad (1)$$

where $\Delta\lambda_B$ denotes $\lambda_B$ shift, $K_\varepsilon$ and $K_T$ represent the strain and temperature sensitivity coefficients, respectively, $\Delta\varepsilon$ is the axial strain applied to the FBG, and $\Delta T$ indicates the temperature variation.

Based on the basic sensing principle of FBG temperature-strain, the sensor measures the strain and temperature at 6 positions on the TL surface, denoted as strain $\varepsilon_{\#i}$ and temperature $T_{\#i}$ ($i$=1~6). Each FBG-$\varepsilon$ in the CL-FBG Sensor corresponds to a FBG-$T$ for temperature compensation of the FBG-$\varepsilon$. Equation (2) represents the temperature measurement model for the 6 position sensing units.

$$T_{\#i} = \Delta\lambda_{T_{\#i}} / K_{T_{T_{\#i}}} + T_{\#i_0} \quad (2)$$

where $T_{\#i}$ denotes the measured temperature, $\Delta\lambda_{T_{\#i}}$ represents the wavelength shift of the FBG-$T$, $K_{TT_{\#i}}$ is the temperature coefficient of the FBG-$T_{\#i}$, and $T_{\#i_0}$ is the initial temperature value.

Equation (3) represents the strain measurement model for the 6 position sensing units.

$$\varepsilon_{\#i} = \Delta\lambda_{\varepsilon_{\#i}} / K_{\varepsilon_{\varepsilon_{\#i}}} - \left(K_{T_{\varepsilon_{\#i}}} \times \Delta T_{\#i}\right)/K_{\varepsilon_{\varepsilon_{\#i}}} + \varepsilon_{\#i_0} \quad (3)$$

where $\varepsilon_{\#i}$ is the strain measurement value, $\Delta\lambda_{\varepsilon_{\#i}}$ is the wavelength change of the FBG-$\varepsilon_{\#i}$, $K_{\varepsilon\varepsilon_{\#i}}$ is the strain coefficient of the FBG-$\varepsilon_{\#i}$. $\Delta T_{\#i}$ is the temperature change value obtained from Equation (2), and $\varepsilon_{\#i_0}$ is the initial strain value.

The deformation of TL under force is not uniaxial but presents a complex two-dimensional distribution in the TL cross-section [28]. During plane strain analysis, a set of 3 strain sensors (strain rosette) arranged at specific orientations can detect the strain state in the plane. The strain transformation equation can be expressed as follows [26]:

$$\begin{cases} \varepsilon_{\alpha_1} = \dfrac{\varepsilon_x + \varepsilon_y}{2} + \dfrac{\varepsilon_x - \varepsilon_y}{2}\cos 2\alpha_1 - \dfrac{\gamma_{xy}}{2}\sin 2\alpha_1 \\ \varepsilon_{\alpha_2} = \dfrac{\varepsilon_x + \varepsilon_y}{2} + \dfrac{\varepsilon_x - \varepsilon_y}{2}\cos 2\alpha_2 - \dfrac{\gamma_{xy}}{2}\sin 2\alpha_2 \\ \varepsilon_{\alpha_3} = \dfrac{\varepsilon_x + \varepsilon_y}{2} + \dfrac{\varepsilon_x - \varepsilon_y}{2}\cos 2\alpha_3 - \dfrac{\gamma_{xy}}{2}\sin 2\alpha_3 \end{cases} \quad (4)$$

where $\varepsilon_x$, $\varepsilon_y$, and $\gamma_{xy}$ represent the normal strains on the X-axis and Y-axis, and the shear strain in the XY plane, respectively.

$\varepsilon_{\alpha_1}$, $\varepsilon_{\alpha_2}$, and $\varepsilon_{\alpha_3}$ are the strain values at the 3 measurement points, while $\alpha_1$, $\alpha_2$, and $\alpha_3$ are the angles between these 3 measurement points and the X-axis.

Based on $\varepsilon_x$, $\varepsilon_y$, and $\gamma_{xy}$, the magnitudes ($\varepsilon_1$, $\varepsilon_2$) and directions ($\theta$) of the principal strains in the plane can be obtained using Equation (5).

$$\begin{cases} \varepsilon_{1,2} = \frac{\varepsilon_x + \varepsilon_y}{2} \pm \sqrt{(\frac{\varepsilon_x - \varepsilon_y}{2})^2 + (\frac{\gamma_{xy}}{2})^2} \\ \tan(2\theta) = \frac{\gamma_{xy}}{\varepsilon_x - \varepsilon_y} \end{cases} \quad (5)$$

The bending of TL and the different shapes of ice both cause uneven cross-sectional deformation. The principal strains measured by a single strain rosette cannot accurately represent the deformation state of TL cross-section. Therefore, the sensing model of the CL-FBG Sensor is proposed, as shown in Fig. 3.

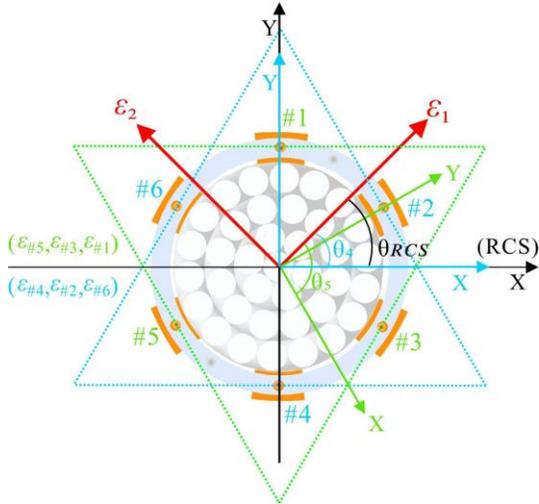

Fig. 3: Theoretical model of plane principal strain for CL-FBG Sensor

A strain rosette consists of 3 measurement points, each located along the sides of an equilateral triangle. By using each strain unit as the first measurement point of the strain rosette, a total of 6 strain rosettes are used to calculate the plane principal strains.

The 3 measurement points of each strain rosette have angles of 0°, 60°, and 120° with respect to the X-axis of the selected coordinate system. The plane principal strain model of the CL-FBG Sensor is as follows:

$$\begin{cases} \varepsilon_1 = \frac{\varepsilon_{\alpha_1} + \varepsilon_{\alpha_2} + \varepsilon_{\alpha_3}}{3} + \frac{\sqrt{(2\varepsilon_{\alpha_1} - \varepsilon_{\alpha_2} - \varepsilon_{\alpha_3})^2 + 3(\varepsilon_{\alpha_3} - \varepsilon_{\alpha_2})^2}}{3} \\ \varepsilon_2 = \frac{\varepsilon_{\alpha_1} + \varepsilon_{\alpha_2} + \varepsilon_{\alpha_3}}{3} - \frac{\sqrt{(2\varepsilon_{\alpha_1} - \varepsilon_{\alpha_2} - \varepsilon_{\alpha_3})^2 + 3(\varepsilon_{\alpha_3} - \varepsilon_{\alpha_2})^2}}{3} \\ \theta = \frac{\arctan\left(\frac{\sqrt{3}(\varepsilon_{\alpha_3} - \varepsilon_{\alpha_2})}{2\varepsilon_{\alpha_1} - \varepsilon_{\alpha_2} - \varepsilon_{\alpha_3}}\right)}{2} \end{cases} \quad (6)$$

When the planar deformation is uniform, the plane principal strains $\varepsilon_1$ and $\varepsilon_2$ obtained from the 6 strain rosettes are equal in magnitude, and the angle $\theta_{RCS}$ between $\varepsilon_1$ and the X-axis of the reference coordinate system satisfies Equation (7).

$$\theta_{RCS} = \theta + 20\sqrt{3} \times (-1)^i \times \sin\left(\frac{\pi(i-1)}{3}\right) \quad (7)$$

## III. FEM OF CL-FBG SENSOR

To simulate the response of the CL-FBG Sensor to the bending of the transmission line, a finite element model (FEM) is created. The TL is simplified as a circular model with a radius of 13 mm and a height of 800 mm. The model parameters for this simulation are listed in Table I. The simulation model is set as a simply supported beam model. Displacement loads are applied at the deployment locations of the sensor to simulate the bending deformation of the TL caused by ice accretion. The direction of the displacement load is from the action location toward the center of the circle. The range of displacement load $d$ (deflection) is from 0 to 2 mm, with a step size of 0.1 mm. The simulation results are shown in Fig. 4.

TABLE I
PHYSICAL PARAMETERS OF SIMULATION MODEL

| Component | Material | Young's modulus/Gpa | Poisson's ratio |
|---|---|---|---|
| TL | Aluminium | 60.00 | 0.30 |
| Optical fiber | Silica | 72.00 | 0.17 |
| CL-FBG Sensor | Plastic | 2.64 | 0.44 |

A displacement load is applied from point $d_0$ in Fig. 4(a) toward the center of the circle, with the direction of the displacement load forming an angle of 0° with the Y-axis. A negative strain value represents compression, while a positive strain value represents tension. The results are shown in Fig. 4(b). Compression deformation occurs at positions #1, #2, and #6 of TL, while stretching deformation occurs at positions #3, #4, and #5. The deformation values at symmetric positions along the X-axis are equal, with the maximum deformation occurring at positions #1 and #4.

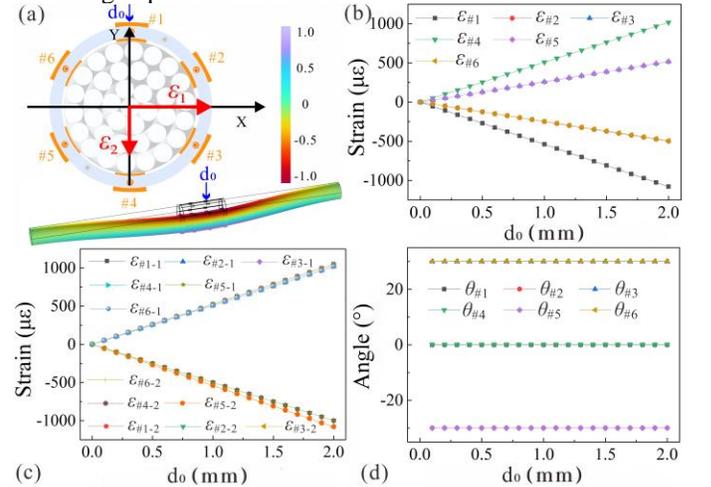

Fig 4. Strain and plane principal strain under $d_0$ load. (a) Simulation experiment. (b) measured strain. (c) Plane principal strain (d) Plane principal strain angle.

As shown in Fig. 4(c~d), the plane principal strains $\varepsilon_1$ and $\varepsilon_2$ obtained from the 6 strain rosettes are equal in magnitude and have the same direction relative to RCS (satisfying Equation (7)). This indicates that the deformation of TL cross-section is uniform. Under the $d_0$ load, the direction of $\varepsilon_1$ is along the positive direction of the X-axis ($\theta_{RCS}=0°$), which is perpendicular to the bending direction of TL. The relationship between the plane principal strain and the bending magnitude of TL can be characterized by the ratio $k$ of $\varepsilon_1$ to deflection. Here, $k = 515.29$ με/mm.

To further verify the bending response of the sensor under loads from different directions, displacement loads $d_\alpha$ are applied at positions with angles $\alpha$ of 10°, 20°, and 30° relative to the Y-axis. The results are shown in Fig. 5.

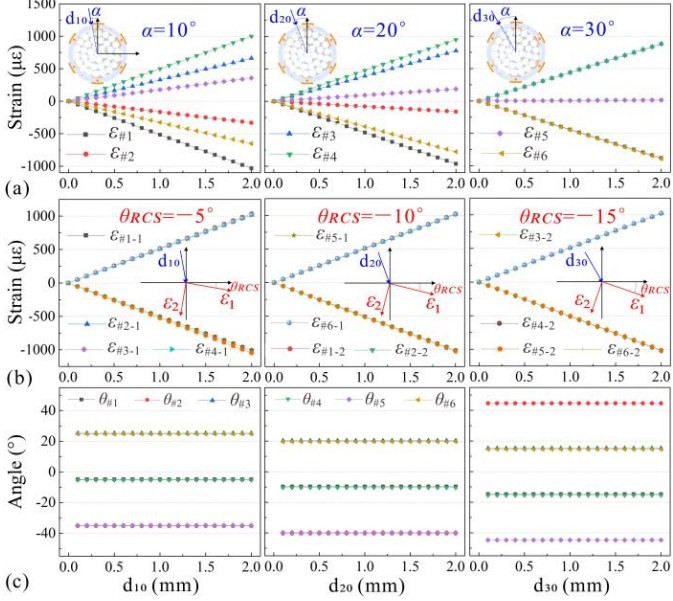

Fig 5. Strain and plane principal strain under $d_{10}$, $d_{20}$, and $d_{30}$ loads. (a) Strain from the 3 experiments. (b) Plane principal strain. (c) Plane principal strain angle.

The strains under $d_{10}$, $d_{20}$, and $d_{30}$ loads are shown in Fig. 5(a). The 6 strain values are symmetric along the X-axis. Under the $d_{30}$ load, there is no deformation at positions #2 and #5, as they are perpendicular to the load direction, so $\varepsilon_{\#2} = \varepsilon_{\#5} = 0$ με. As shown in Fig. 5(b~c), under the same load, the $\varepsilon_1$ and $\varepsilon_2$ obtained from the 6 strain rosettes are equal in magnitude. The $\theta$ satisfies Equation (7), indicating that the directions of $\varepsilon_1$ relative to RCS are the same. Based on the above simulation results, both the direction and magnitude of the $\varepsilon_1$ and the $d$ satisfy Equation (8). This proves that the CL-FBG Sensor can measure both the magnitude and direction of the bending deformation of TL.

$$\begin{cases} \alpha = -\left(\theta_{RCS}/2 + \sum_{i=1}^{6}\theta_{\#i}/4\right) \\ d = \varepsilon_1 / k, \ (k=515.29 \ \mu\varepsilon/mm) \end{cases} \quad (8)$$

where $\alpha$ represents the angle between the bending deformation and the Y-axis., $\theta_{\#i}$ denotes the angles calculated by the 6 strain rosettes, and d is the deflection.

## IV. EXPERIMENTS AND RESULTS

### A. Calibration experiment

To obtain the sensing model of the CL-FBG Sensor, a calibration experiment is designed, as shown in Fig. 6(a). In the temperature calibration experiment, the sensor is placed inside a incubator. A Fluke-5609 platinum resistance thermometer (with an accuracy of ±0.008°C) is positioned near the sensor. The platinum resistance thermometer records the temperature using a Fluke-2638A data acquisition instrument, and its readings are considered the reference temperature values in the experiment. The temperature range for the temperature calibration experiment is from -30°C to 10°C, with a step size of 10°C. After allowing the temperature to stabilize at each point, the $\lambda_B$ of each sensor unit is recorded using a demodulator.

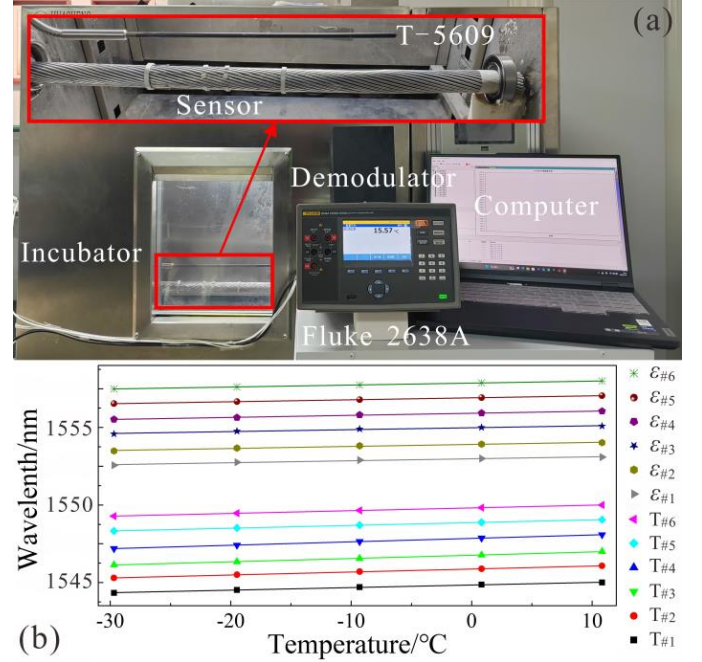

Fig 6. Temperature calibration experiment. (a) Experimental platform. (b) Calibration results.

The temperature calibration experiment was repeated 3 times, yielding the temperature coefficients for each sensor unit. The results are shown in Fig. 6(b). The $\lambda_B$ of the 6 FBG-$\varepsilon$ and 6 FBG-$T$ exhibit a good linear relationship with temperature. The temperature coefficients of the 6 FBG-$\varepsilon$ range from 12.64 to 13.32 pm/°C, with a linearity greater than 0.996. The temperature coefficients of the 6 FBG-$T$ range from 16.58 to 22.10 pm/°C, with a linearity greater than 0.997, as shown in Table II. The strain coefficient of the FBG is 1.19 pm/με, with an average error of 1.418% and an average resolution of 2.251%. The sensor strain range is from -2000 to 2000 με [29].

TABLE II
TEMPERATURE COEFFICIENT OF SENSING UNIT

|  | $\varepsilon_{\#1}$ | $\varepsilon_{\#2}$ | $\varepsilon_{\#3}$ | $\varepsilon_{\#4}$ | $\varepsilon_{\#5}$ | $\varepsilon_{\#6}$ |
|---|---|---|---|---|---|---|
| $R^2$ | 0.998 | 0.996 | 0.998 | 0.997 | 0.999 | 0.998 |
| $K_T$ /(pm/°C) | 12.64 | 12.93 | 12.44 | 13.32 | 12.56 | 12.71 |
|  | $T_{\#1}$ | $T_{\#2}$ | $T_{\#3}$ | $T_{\#4}$ | $T_{\#5}$ | $T_{\#6}$ |
| $R^2$ | 0.997 | 0.999 | 0.999 | 0.999 | 0.998 | 0.999 |
| $K_T$ /(pm/°C) | 16.58 | 19.14 | 21.21 | 22.10 | 17.96 | 17.76 |

The $K_T$ and $K_\varepsilon$ from Equations (2) and (3) have been obtained. Therefore, the sensing model for the CL-FBG Sensor is as follows:

$$\begin{bmatrix} T_{\#1} \\ \vdots \\ T_{\#6} \\ \varepsilon_{\#1} \\ \vdots \\ \varepsilon_{\#6} \end{bmatrix} = \begin{bmatrix} \frac{1}{16.58} & 0 & 0 & 0 & 0 & 0 \\ \vdots & \vdots & \vdots & \vdots & \vdots & \vdots \\ 0 & 0 & 0 & 0 & 0 & \frac{1}{17.76} \\ \frac{1}{12.64} & 0 & 0 & 0 & 0 & 0 \\ \vdots & \vdots & \vdots & \vdots & \vdots & \vdots \\ 0 & 0 & 0 & 0 & 0 & \frac{1}{17.27} \end{bmatrix} \begin{bmatrix} \Delta\lambda_{T_{\#1}} \\ \vdots \\ \Delta\lambda_{T_{\#6}} \\ \Delta\lambda_{\varepsilon 1} \\ \vdots \\ \Delta\lambda_{\varepsilon 6} \end{bmatrix} + \begin{bmatrix} T_{\#1_0} \\ \vdots \\ T_{\#6_0} \\ \varepsilon_{\#1_0} \\ \vdots \\ \varepsilon_{\#6_0} \end{bmatrix} - \begin{bmatrix} 0 \\ \vdots \\ 0 \\ \frac{16.64\Delta T_{\#1}}{1.19} \\ \vdots \\ \frac{17.27\Delta T_{\#6}}{1.19} \end{bmatrix} \quad (9)$$

### B. Weight load experiment

The experiment further simulates the ice load on the TL using a weighted object, as shown in Fig. 7. During the experiment, the position #1 of TL is kept at the top. A bracket is designed to protect the optical fiber. To simulate uniform axial ice loading, weights are hung at 3 positions, spaced 15 cm apart from the sensor. The mass at each position ranges from 0 to 500 g, with a step size of 50 g.

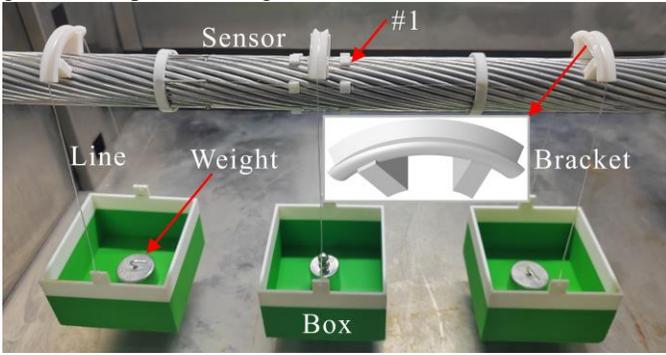

Fig 7. Weight load experimental setup.

As shown in Fig. 8(a), $\varepsilon_{\#1}$, $\varepsilon_{\#2}$, and $\varepsilon_{\#6}$ are negative, indicating that the upper half of TL is compressed; $\varepsilon_{\#3}$, $\varepsilon_{\#4}$, and $\varepsilon_{\#5}$ are positive, indicating that the upper half of TL is stretched. The actual results are consistent with the simulation results. When the mass of the mounted weight is 1500 g, the maximum compression deformation occurs at position $\varepsilon_{\#1}$, with a strain of -382.7 με, and the maximum tension deformation occurs at position $\varepsilon_{\#4}$, with a strain of 391.7 με. The data in Fig. 8(a) is fitted with a second-degree curve, and the parameters of the fitting curve are shown in Table III. The relationship between the strain $\varepsilon_{\#i}$ at the 6 positions of TL and the ice mass M is given by Equation (10).

$$\varepsilon_{\#i} = a_{\#i}M^2 + b_{\#i}M + c_{\#i} \quad (10)$$

The magnitudes of the $\varepsilon_1$ and $\varepsilon_2$ increase as the mounted weight increases, with the maximum values reaching 390.4 με and -383.3 με, respectively. The $\varepsilon_1$ and $\varepsilon_2$ measured by the 3 strain rosettes at points #1, #3, and #5 are equal, and similarly, the measurements at points #2, #4, and #6 are also identical, as shown in Fig. 8(b). The angles $\theta_{\#1}$ and $\theta_{\#4}$, $\theta_{\#2}$ and $\theta_{\#5}$, and $\theta_{\#3}$ and $\theta_{\#6}$ are essentially equal, with values of 0°, -30°, and 30°, respectively, as shown in Fig. 8(c). Since the direction of gravity is vertically downward, it is consistent with the results of applying a 0° displacement load in the simulation. The angle $\beta$ between $\varepsilon_1$ and the bending is 90°, which verifies the applicability of Equation (8).

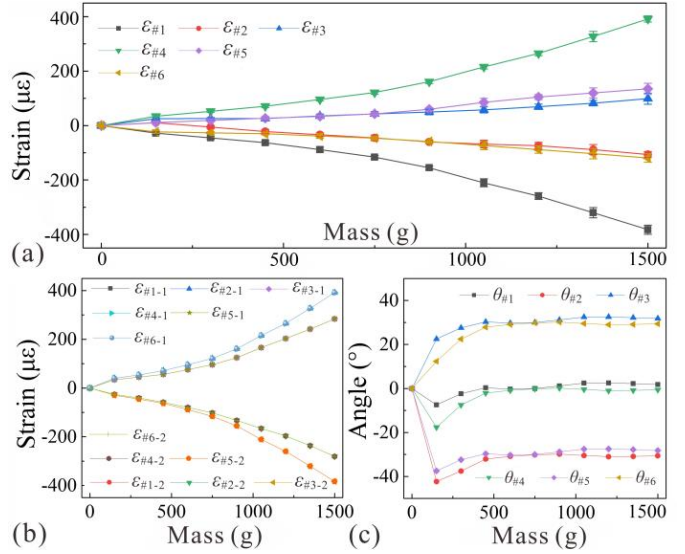

Fig 8: Results of the weight load experiment. (a) Strain measured by the Sensor. (b) Plane principal strain. (c) Plane principal strain angle.

TABLE III
PARAMETERS FOR FITTING CURVES

| Parameter | Value | | | | | |
|---|---|---|---|---|---|---|
| $a_{\#i}$ | -1.31×10⁻⁴ | -7.55×10⁻⁶ | 1.74×10⁻⁵ | 1.29×10⁻⁴ | 3.83×10⁻⁵ | -2.42×10⁻⁵ |
| $b_{\#i}$ | -0.05 | -0.06 | 0.03 | 0.06 | 0.03 | -0.04 |
| $c_{\#i}$ | -10.32 | 9.41 | 10.14 | 13.76 | 2.31 | -8.57 |
| $R^2$ | 0.997 | 0.996 | 0.997 | 0.996 | 0.998 | 0.997 |

### C. Icing experiment on the TL

The design of TL icing experiment is shown, with the experimental setup depicted in Fig. 9(a). Supercooled water droplets are sprayed onto the surface of TL using a blow head in the incubator to simulate ice formation. The required low-temperature environment is provided by the incubator. Electrical de-icing is simulated by embedding a heating wire in the TL. The experiment is divided into 4 stages: the initial stage ($S_0$), freeze stage ($S_1$), stable ice stage ($S_2$), and thaw stage ($S_3$). Specifically, $S_0$ represents the no-ice state of TL; $S_1$ is the spraying and icing stage; $S_2$ is the stage of changing the ambient temperature after the completion of icing.; and $S_3$ implements electrothermal de-icing via Joule heating. The ambient temperature for $S_0$ is set to -20°C; for S1, the ambient temperature remains unchanged; for $S_2$, the ambient temperature is adjusted to -16°C, -11°C, and -5°C; in $S_3$, the heating wire power is set to 1W/cm. Heating is stopped once the ice has fully melted, and the ambient temperature is maintained until TL returns to the $S_0$ state.

According to actual operating conditions, during the $S_1$ stage: TL is kept stationary to simulate the formation of C-shaped ice (with a thickness of 10.1 mm) [30], as shown in Fig. 9(b); TL is rotated to simulate the formation of circular ice under wind load (with a thickness of 10.2 mm) [5], as shown in Fig. 9(c). Experimental data is recorded every 20 s, and the duration of each stage is shown in Table IV.

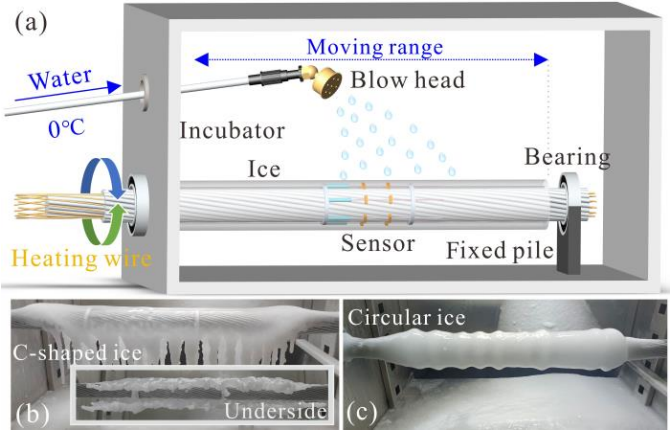

Fig 9. Experiment on detecting the freeze-thaw process of ice accretion. (a) Ice accretion experimental setup. (b) C-shaped ice experiment. (c) Circular ice experiment.

TABLE IV
DURATION OF EACH STAGE IN THE ICING EXPERIMENT

| Ice shape | S0/s | S1/s | S2/s | S3/s | S0/s |
|---|---|---|---|---|---|
| C-shaped ice | 2100 | 3680 | 23920 | 8880 | 2460 |
| Circular ice | 2400 | 6360 | 22860 | 7500 | 3120 |

First, analyzing the C-shaped ice experiment, as shown in Fig. 10. In the $S_0$ stage, there is no ice. The $\varepsilon_{\#1} \sim \varepsilon_{\#6}$ are all 0 με. The $T_{\#1} \sim T_{\#6}$ are all consistent with the ambient temperature (-20°C).

In the $S_1$ stage, when the supercooled water comes into contact with the low-temperature TL, a phase change occurs. This phase change releases heat and is accompanied by a volume expansion. As a result, both strain and temperature increase suddenly. Afterward, the temperature stabilizes at the phase transition temperature of water(0°C).

In the $S_2$ stage, once the icing is complete, the ambient temperature is changed. During this phase, the $\varepsilon_{\#1}$, $\varepsilon_{\#2}$, $\varepsilon_{\#3}$, $\varepsilon_{\#5}$, and $\varepsilon_{\#6}$ undergo thermal expansion deformation as the temperature changes. Since point #4 is located at the bottom of TL and is not covered by ice (Fig. 9(b)), $\varepsilon_{\#4}$ does not exhibit significant changes with temperature and remains almost a straight line. At this stage, the average strain of $\varepsilon_{\#4}$ is 105.68 με. Based on the relationship between $\varepsilon_{\#4}$ and ice mass in Equation (10), the mass of the C-shaped ice is determined to be 650.76 g. After the experiment, the mass of the melted water is measured to be 602.43 g, resulting in an error of 48.33 g (8.02%).

In the $S_3$ stage, after heating is activated, the C-shaped ice reaches the phase transition temperature of 0°C after 380 s and maintains this temperature for 440 s. The strain value reaches its maximum when the phase transition temperature is first reached, and then decreases as the ice melts, as shown in the enlarged detail of the $T_{\#2}$ and $\varepsilon_{\#2}$ curves in Figure 9. Since the ice is in the phase transition state, the melted ice continues to come into contact and collide with TL. Additionally, the water from the melted ice flows along the TL and over the sensor, causing the temperature and strain readings to become erratic and chaotic during this phase.

When $T_{\#1} \sim T_{\#6}$ all rise and $\varepsilon_{\#1} \sim \varepsilon_{\#6}$ show no significant fluctuations, it indicates that the ice has completely detached from TL. At this point, heating is stopped, and the ambient temperature is maintained until TL and CL-FBG Sensor return to the initial state ($S_0$).

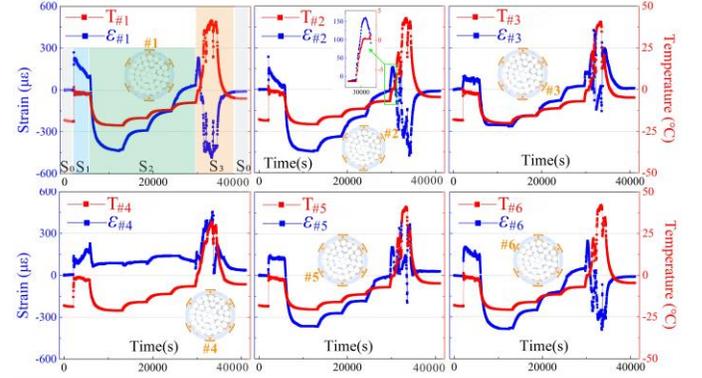

Fig 10. Strain and temperature in the C-shaped ice experiment.

The analysis of the circular ice experiment is shown in Fig. 11. In the $S_0$ and $S_1$ stages, the strain and temperature of the circular ice exhibit similar patterns to those of the C-shaped ice. However, in the $S_2$ stage, the surface of TL freezes uniformly (Fig. 9(c)), and $\varepsilon_{\#4}$ undergoes thermal expansion as the temperature changes. In the $S_3$ stage, under the conditions of equal ambient temperature, ice thickness, and heating power, the melting durations for circular ice and C-shaped ice are 7500 s and 8880 s, respectively. The melting duration for C-shaped ice is 18.4% longer than that for circular ice, indicating that the ice shape significantly affects the de-icing duration.

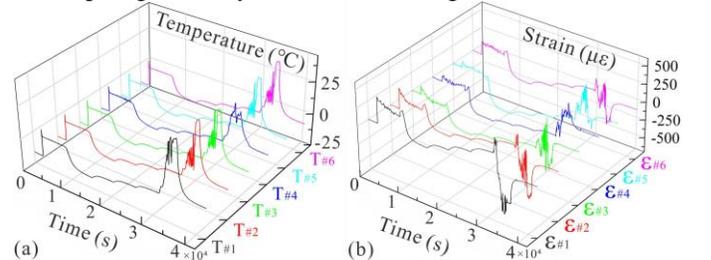

Fig 11. Experimental results of circular ice. (a) Strain at 6 locations. (b) Temperature at 6 locations.

Based on the strain measured by the CL-FBG Sensor and sensing model, the plane principal strain for both ice shape were obtained, as shown in Fig. 12. In the $S_0$ stage, the $\varepsilon_1$ is 0 με. In the S1 stage, the maximum values of $\varepsilon_1$ in the C-shaped ice and circular ice experiments are 270.3 με and 356.3 με, respectively.

In the $S_2$ stage, in the C-shaped ice experiment, $\varepsilon_{\#2-1}$, $\varepsilon_{\#4-1}$, and $\varepsilon_{\#6-1}$ do not change with temperature and appear almost as straight lines, indicating that there is no ice at position #4 of TL. However, in the circular ice experiment, $\varepsilon_{\#2-1}$, $\varepsilon_{\#4-1}$, and $\varepsilon_{\#6-1}$ show a distinct stepped pattern as the temperature changes. Based on the differences in principal strains, it is possible to determine whether the ice shape is circular or C-shaped, as shown in Fig. 12(a~b).

In the $S_3$ stage, the ice accretion is unstable due to the phase transition process, causing continuous fluctuations in the magnitude of the plane principal strains. After heating is stopped, both TL and the sensor return to initial state ($S_0$).

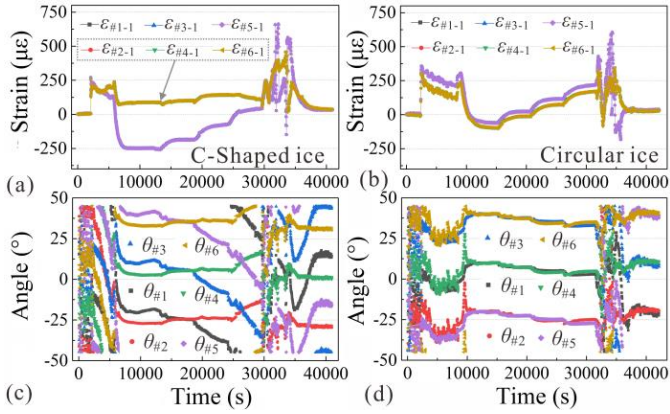

Fig 12. Plane principal strains in 2 icing experiments. (a) $\varepsilon_1$ in the C-shaped ice experiment. (b) $\varepsilon_1$ in the circular ice experiment. (c) $\theta$ in the C-shaped ice experiment. (d) $\theta$ in the circular ice experiment

In the $S_1$ and $S_3$ stages, due to the phase transition between ice and water, the internal state is unstable, causing significant fluctuations in $\theta$ during these periods. In the S2 stage, $\theta$ remains relatively stable ($\theta_{RCS} \approx 0°$). The ice accretion exerts a vertically downward gravitational force on the TL, causing TL to bend downward. The sum of $\theta_{\#1} \sim \theta_{\#6}$ is $0°$, leading to the conclusion that the direction of TL bending coincides with the Y-axis (vertically downward), thus satisfying Equation (8). This is also consistent with the simulation experiment under the $d_0$ load.

### D. Judgment of the icing state

The protection and management of ice accretion on the TL require real-time monitoring of the ice formation status. Based on the measurement data, a model for determining the freeze-thaw state has been established, as shown in Fig. 13(a).

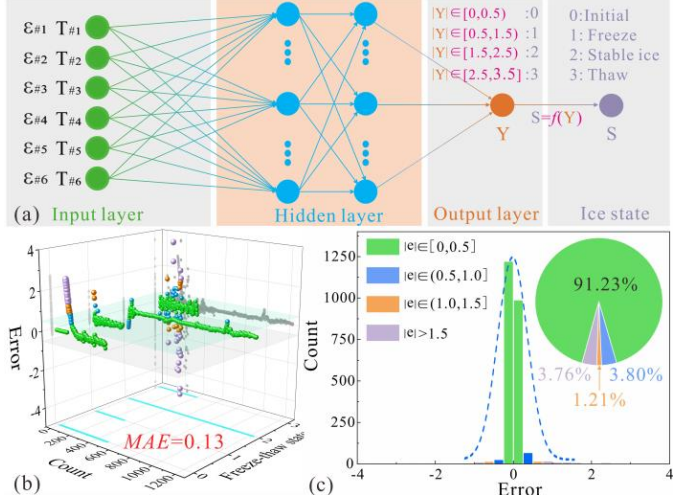

Fig 13. BP neural network for determining freeze-thaw states. (a) BP neural network. (b) Judgment error. (c) Error distribution.

The BP neural network structure is a 3 layer network [31]. The dataset includes 6 independent icing experiments, with 3 C-shaped ice experiments and 3 circular ice experiments. The strain and temperature measurements obtained from the experiments are combined with the corresponding icing states to form the neural network dataset. This dataset will serve as the input for the BP neural network model to predict and analyze the freeze-thaw process of ice accretion. 5 experiments are randomly selected from the dataset as the training set (11,979 data points), and the remaining one is used as the validation set (2,419 data points).

In the model, the optimal number of hidden layer nodes is set to 262, the number of training iterations is set to 10,000, the learning rate is 0.01, and the minimum error goal for training is set to $10^{-7}$. After training, the model's mean absolute error (MAE) is 0.13. The error in the icing state predicted by the model is shown in Fig. 13(b~c). If the absolute error $|e|$ is set to be less than 0.5, indicating correct identification of the ice accretion state, the BP neural network achieves an accuracy of 91.23% in identifying the icing state of TL.

### E. Analysis of the ice accretion freeze-thaw process on transmission lines.

The variation of the freeze-thaw states of ice accretion on the TL can be represented in Fig. 14. The $S_0$ stage is the initial state. Under low-temperature conditions, supercooled water come into contact with TL and begin to freeze. At this point, TL enters the $S_1$ stage. Once the ice stabilizes and enters the $S_2$ stage, the ice shape on the TL (C-shaped ice or circular ice) can be determined by analyzing the plane principal strain curve. Once the system enters the $S_3$ stage, the CL-FBG Sensor can support de-icing strategies, such as determining the time for electric heating de-icing and the DC current required. When the CL-FBG Sensor detects that the de-icing is complete, the electric heating de-icing is immediately turned off, and the TL returns to the initial state $S_0$. Therefore, by monitoring the ice shape and freeze-thaw states, the scientific decision-making capability for TL ice prevention and control can be effectively improved.

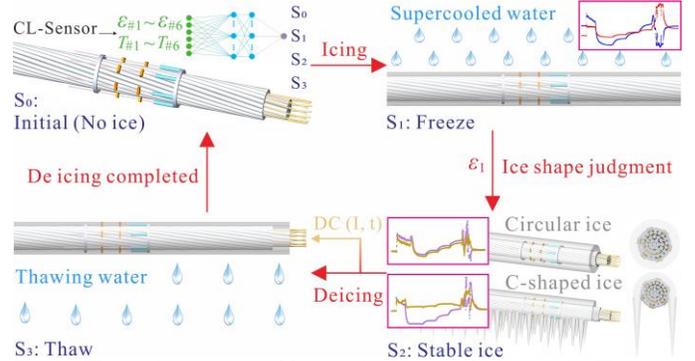

Fig 14. The entire process of freeze-thaw state of icing on the TL

### V. CONCLUSION

In response to the need for deformation monitoring throughout the freeze-thaw process of ice accretion on the TL, the CL-FBG Sensor based on the principle of plane principal strain is proposed. Through FEM analysis under different bending conditions, it is demonstrated that the plane principal strain measured by the sensor can characterize the bending condition of the TL in any direction of the cross-section. Further calibration of the sensor was conducted, with a measurement range of -2000 to 2000 με, an average error of 1.418%, and an average resolution of 2.251%. An ice accretion freeze-thaw experimental platform was built to investigate the relationship between ice mass and bending strain, verifying the simulation results. Through icing experiments, it was found that the ice shape on the TL cross-section (C-shaped ice and circular ice) can be identified based on the distribution of plane principal strain. A BP neural network model was developed to determine the four states (no ice/freeze/stable/thaw) during the icing process on TL, with an accuracy of 91.23% and the MAE of 0.13. This study can detect the bending and ice accretion load of the TL, as well as make high-precision judgment on ice shape and freeze-thaw state, providing a new method for optimizing de-icing strategies for TL.

## REFERENCES

[1] S. Zhang, and W. Chen, "Assessing the energy transition in China towards carbon neutrality with a probabilistic framework," *Nature Communications*, vol. 13, no. 1, 2022.
[2] X. Li, W. Yang, Y. Liao et al. "Short-term risk-management for hydro-wind-solar hybrid energy system considering hydropower part-load operating characteristics," *Applied Energy*, vol. 360, 2024.
[3] R. Zhou, Z. Zhang, H. Zhang et al., "Reliable monitoring and prediction method for transmission lines based on FBG and LSTM," *Advanced Engineering Informatics*, vol. 62, 2024.
[4] C. Yang, H. Su, H. Ji et al., "Multifunctional laser-ablated lubricant-infused slippery film with self-sensing and anti-icing/deicing properties," *Surfaces and Interfaces*, vol. 55, 2024.
[5] G. Yang, X. Jiang, Y. Liao et al., "Research on load transfer melt-icing technology of transmission lines: Its critical melt-icing thickness and experimental validation," *Electric Power Systems Research*, vol. 221, 2023.
[6] Y. Huang, Y. Chen, X. Yang et al., "Study on icing characteristics of bundled conductor of transmission line based on shadowing effect analysis," *Cold Regions Science and Technology*, vol. 231, 2025.
[7] X. Huang, F. Zhang, H. Li, and X. Liu, "An Online Technology for Measuring Icing Shape on Conductor Based on Vision and Force Sensors," *IEEE Transactions on Instrumentation and Measurement*, vol. 66, no. 12, pp. 3180-3189, 2017.
[8] B. Li, G. Thomas, and D. Williams, "Detection of Ice on Power Cables Based on Image Texture Features," *IEEE Transactions on Instrumentation and Measurement*, vol. 67, no. 3, pp. 497-504, 2018.
[9] L. Yang, Z. Chen, Y. Hao et al., "Icing detection method of 500 kV overhead ground wire with tangent tower under special terrain based on vertical tension measurement technology," *Cold Regions Science and Technology*, vol. 210, 2023.
[10] L. Yang, Y. Chen, Y. Hao et al., "Detection Method for Equivalent Ice Thickness of 500-kV Overhead Lines Based on Axial Tension Measurement and Its Application," *IEEE Transactions on Instrumentation and Measurement*, vol. 72, pp. 1-11, 2023.
[11] X. Jiang, Z. Xiang, Z. Zhang et al., "Predictive Model for Equivalent Ice Thickness Load on Overhead Transmission Lines Based on Measured Insulator String Deviations," *IEEE Transactions on Power Delivery*, vol. 29, no. 4, pp. 1659-1665, 2014.
[12] G. Yang, Y. Liao, X. Jiang et al., "Research on Value-Seeking Calculation Method of Icing Environmental Parameters Based on Four Rotating Cylinders Array," *Energies*, vol. 15, no. 19, 2022.
[13] M. J. Moser, T. Bretterklieber, H. Zangl, and G. Brasseur, "Strong and Weak Electric Field Interfering: Capacitive Icing Detection and Capacitive Energy Harvesting on a 220-kV High-Voltage Overhead Power Line," *IEEE Transactions on Industrial Electronics*, vol. 58, no. 7, pp. 2597-2604, 2011.
[14] L. Chen, H. Zhang, Q. Wu, and V. Terzija, "A Numerical Approach for Hybrid Simulation of Power System Dynamics Considering Extreme Icing Events," *IEEE Transactions on Smart Grid*, vol. 9, no. 5, pp. 5038-5046, 2018.
[15] L. Yang, J. Chen, Y. Hao et al., "Experimental Study on Ultrasonic Detection Method of Ice Thickness for 10 kV Overhead Transmission Lines," *IEEE Transactions on Instrumentation and Measurement*, vol. 72, pp. 1-10, 2023.
[16] R. Zhou, Z. Zhang, T. Zhai et al., "Machine learning-based ice detection approach for power transmission lines by utilizing FBG micro-meteorological sensors," *Optics Express*, vol. 31, no. 3, 2023.
[17] H. Wang, G. Xiao, L. Wang et al., "A Novel Approach to Partial Discharge Detection Under Repetitive Unipolar Impulsive Voltage," *IEEE Transactions on Industrial Electronics*, vol. 70, no. 11, pp. 11681-11691, 2023.
[18] J. Sun, Z. Zhang, Y. Li et al., "Distributed Transmission Line Ice-Coating Recognition System Based on BOTDR Temperature Monitoring," *Journal of Lightwave Technology*, vol. 39, no. 12, pp. 3967-3973, 2021.
[19] Z. Xu, S. Song, L. Zhao, and X. Li, "OPGW Icing Monitoring Method Based on Phase Difference Between Temperature Curves," *IEEE Transactions on Power Delivery*, vol. 39, no. 2, pp. 1303-1306, 2024.
[20] Z.-W. Ding, X.-P. Zhang, N.-M. Zou et al., "Phi-OTDR Based On-Line Monitoring of Overhead Power Transmission Line," *Journal of Lightwave Technology*, vol. 39, no. 15, pp. 5163-5169, 2021.
[21] S. Fan, X. Jiang, L. Shu et al., "DC Ice-Melting Model for Elliptic Glaze Iced Conductor," *IEEE Transactions on Power Delivery*, vol. 26, no. 4, pp. 2697-2704, 2011.
[22] Y. Luo, C. Gao, D. Wang et al., "Predictive model for sag and load on overhead transmission lines based on local deformation of transmission lines," *Electric Power Systems Research*, vol. 214, 2023.
[23] M. Zhang, Y. Xing, Z. Zhang, and Q. Chen, "Design and Experiment of FBG-Based Icing Monitoring on Overhead Transmission Lines with an Improvement Trial for Windy Weather," *Sensors*, vol. 14, no. 12, pp. 23954-23969, 2014.
[24] G. Ma, N. Mao, Y. Li et al., "The Reusable Load Cell with Protection Applied for Online Monitoring of Overhead Transmission Lines Based on Fiber Bragg Grating," *Sensors*, vol. 16, no. 6, 2016.
[25] R. Zhou, Z. Zhang, Z. Yan et al., "GPR-based high-precision passive-support fiber ice coating detection method for power transmission lines," *Optics Express*, vol. 29, no. 19, 2021.
[26] V. Biazi-Neto, C. A. F. Marques, A. Frizera-Neto, and A. G. Leal-Junior, "FBG-Embedded Robotic Manipulator Tool for Structural Integrity Monitoring From Critical Strain-Stress Pair Estimation," *IEEE Sensors Journal*, vol. 22, no. 6, pp. 5695-5702, 2022.
[27] B. Jia, X. Deng, Z. Qin et al., "Six-Dimensional Strain Sensor Based on Fiber Bragg Gratings for Frozen Soil," *IEEE Transactions on Instrumentation and Measurement*, vol. 73, pp. 1-10, 2024.
[28] L. Xiang, H. Y. Wang, Y. Chen et al., "Elastic-plastic modeling of metallic strands and wire ropes under axial tension and torsion loads," *International Journal of Solids and Structures*, vol. 129, pp. 103-118, 2017.
[29] B. Jia, C. Du, X. Deng et al., "FBG-LPFG-Based Sensor to Monitor 3-D Strain in Ice During Freezing–Melting Processes," *IEEE Sensors Journal*, vol. 23, no. 9, pp. 9333-9342, 2023.
[30] X. Han, P. Sun, B. Xing et al., "Influence of torsion on icing process of transmission lines," *IET Generation, Transmission & Distribution*, vol. 16, no. 20, pp. 4230-4238, 2022.
[31] B. Jia, X. Deng, C. Du, et al., "Wide Wavelength Range Demodulation Method of Cascaded FBGs Based on LPFG Spectral Control," *Journal of Lightwave Technology*, vol. 42, no. 17, pp. 6090-6098, 2024.